# Effects of detection-beam focal offset on displacement detection in optical tweezers

ANNI CHEN,[1] HUI LUO,[1] ZHIJIE CHEN,[1] HANNING FENG,[1] TENGFANG KUANG,[1] HUI AN,[1] XIANG HAN,[1,2] WEI XIONG,[1,3] AND GUANGZONG XIAO[1]

[1]*College of Advanced Interdisciplinary Studies, National University of Defense Technology, Changsha Hunan, 410073, China*
[2] *hanxiang05@nudt.edu.cn*
[3] *xiongwei08@nudt.edu.cn*

**Abstract:** A high-resolution displacement detection can be achieved by analyzing the scattered light of the trapping beams from the particle in optical tweezers. In some applications where trapping and displacement detection need to be separated, a detection beam can be introduced for independent displacement detection. However, the detection beam focus possibly deviates from the centre of the particle, which will affect the performance of the displacement detection. In this paper, we detect the radial displacement of the particle by utilizing the forward scattered light of the detection beam from the particle. The effects of the lateral and axial offsets between the detection beam focus and the particle centre on the displacement detection are analyzed by the simulation and experiment. The results show that the lateral offsets will decrease the detection sensitivity and linear range and aggravate the crosstalk between the *x*-direction signal and *y*-direction signal of QPD. The axial offsets also affect the detection sensitivity, an optimal axial offset can improve the sensitivity of the displacement detection substantially. In addition, the influence of system parameters, such as particle radius *a*, numerical aperture of the condenser $NA_c$ and numerical aperture of the objective $NA_o$ on the optimal axial offset are discussed. A combination of conventional optical tweezers instrument and a detection beam provides a more flexible working point, allowing for the active modulation of the sensitivity and linear range of the displacement detection. This work would be of great interest for improving the accuracy of the displacement and force detection performed by the optical tweezers.

## 1. Introduction

Since the first introduction of optical tweezers by Ashkin and his colleagues [1], this indispensable and noninvasive technology has been found to become a powerful tool in biology [2-5], fundamental physics [6,7], and precision measurements [8-11]. Specifically, an optically trapped particle can act as a photonic force probe with high quality factor and low dissipation, enabling precision measurements of many physical quantities such as ultra-weak forces [12,13], accelerations [14], microscopic mass [15], and so on.

The precise detection of the particle's displacement plays an important role in precision measurements [16,17]. In optical tweezers, conventional displacement detection methods include the video-based displacement detection and the laser-based displacement detection. The video-based displacement detection method analyzes the images acquired by a digital camera to obtain the information of the particle's displacements [18-20]. The laser-based displacement detection method usually uses position sensitive detectors [21] or balanced detectors with D-shape mirrors [22,23] to monitor the shifts of scattered light spots, This approach can provide a high spatial and temporal resolution [24-27].

In the conventional laser-based displacement detection methods, the particle's displacement can be deduced by the measurement of the scattered light of the trapping beams from the particle [28]. However, in some applications where the two functions of trapping and displacement detection need to be separated, for example, in experiments where the intensity or location of the trapping beams need to be modulated to control the particle's movement [26,29]. Additionally, in a dual-beam fiber-optic trap system, the forward scattered (FS) light or the back scattered (BS) light of the trapping beams are collected by the trapping fiber end, but the FS light and BS light are difficult to be used for the displacement detection [27,30]. Hence, under these cases, it is difficult to achieve displacement detection using the scattered light from the trapping beams, a detection beam can be introduced for displacement detection. In this situation, the detection beam focus possibly deviates from the particle centre due to the installation errors of the optical elements. Chang et al. introduced a detection laser and investigated the sensitivity and the tracking range as a function of the focal offset between the trapping and the detection beams [31]. Sugino et al. investigated the laser interferometry sensitivity and the feedback stability as a function of particle deviates from the detection laser focus [32]. The aforementioned two studies only discussed the influence of axial offsets between the detection beam focus and particle, but did not account for the lateral offsets.

In this paper, a detection beam with wavelength different from the trapping beams is introduced into our system for a better performance and flexibility of displacement detection. The effects of the lateral and axial offsets on the displacement detection are discussed both by theory and experiment. After the optimization of the offsets, a high sensitivity and low crosstalk of the displacement detection is expected.

## 2. Fundamentals

The schematic of the displacement detection system using a detection beam is shown in Fig. 1. A detection beam with a wavelength of 642 nm is coupled into a single-mode fiber to improve its pointing stability and beam quality. The output beam of the fiber is collimated by a fiber collimator and then expanded to slightly overfill the back aperture of the objective (40×, NA = 0.65). The silica particle is positioned and allowed to adsorb on the coverslip which is mounted on a 3-axis piezo translation stage (Physik Instruments, P-733.3CD) with nanometer resolution. The piezo translation stage provides the ability to move the particle and finely tune the offsets between the detection beam focus and the particle. The objective focuses the expanded detection beam onto the particle. The forward scattered light of the detection beam from the particle are collected by a condenser (40×, NA = 0.65), and then transmitted to a quadrant photodiode (QPD) which locates on a plane conjugated to the back-focal-plane (BFP) of the condenser. The radial displacement of the particle can be deduced from the BFP field distribution formed by the scattered light. This displacement detection method is referred to as the back focal plane interferometry [21].

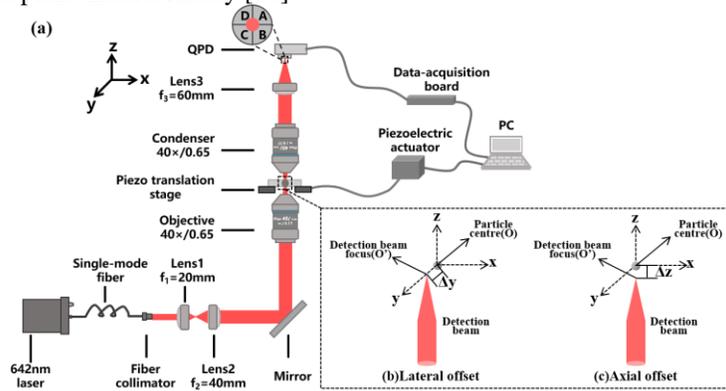

Fig. 1. (a) Schematic of the displacement detection system using a detection beam.
Inset: (b) The diagram of the lateral offset $\Delta y$. (c) The diagram of the axial offset $\Delta z$.

As shown in Fig. 1, the initial position of the particle is defined as the origin (point O, $x_0 = y_0 = z_0 = 0$), the $z$-axis is defined as the propagation direction of the detection beam. Due to the installation error of the optical elements, the detection beam focus possibly deviates from the particle centre, as shown in the inset (b) and (c) of Fig. 1. The lateral offset $\Delta y$ is the distance of the detection beam focus deviating from the particle centre along *y*-axis, and the axial offset $\Delta z$ is the distance of the detection beam focus deviating from the particle centre along *z*-axis.

Numerical calculations based on the T-matrix method are conducted in order to predict the dependence of the sensitivity and the signal crosstalk on the focal offsets. The T-matrix method is applicable for the condition where repeated calculations at different positions are needed for the same particle, greatly improving the efficiency of the simulation [33]. Using the software toolbox—Optical Tweezers Software [34], calculations are conducted using the parameters relevant for our experimental conditions.

## 3. Simulation results

We consider a silica particle with 5-$\mu$m-raduis, this size of the particles is commonly used for the measurements of accelerations in optical tweezers [35,36]. The particle is illuminated by a focused Gaussian beam with wavelength of 642 nm. The refractive index of the particle and medium are 1.46 and 1, respectively. Both the numerical aperture of the objective and condenser is 0.65, and the back aperture of the objective and condenser is 2.9 mm. The power of the detection beam is 1 mW and will not affect the trapping state of the particle in optical tweezers. The axial displacement along the detection beam propagation direction usually can be obtained by using the diaphragm and calculating the total intensity of the scattered light. But the radial displacement along direction perpendicular to the propagation direction of the detection beam is the focus of this study, and it can be obtained from the intensity difference of the scattered light. The focused spot of the detection beam is a symmetric Gaussian spot, when the detection beam focus deviates from the particle centre along *x*-axis or *y*-axis, the response of radial displacement detection is the same, so we only analyze the influence of *y*-axis offset $\Delta y$ and *z*-axis offset $\Delta z$ on the radial displacement detection.

Here, we compare the response curves of the *x*-direction signal and *y*-direction signal of QPD with *x*-axis displacement of the particle under various lateral offsets $\Delta y$, the results were shown in Fig. 2(a) and (b). To quantify the relationship between the focal offset and the performance of the displacement detection, the sensitivity and linear range are introduced and denoted respectively as:

*Sensitivity*: the slope of the response curve of the displacement detection at $x = 0$.

*Linear range*: a central linear region around $x = 0$ where the QPD signal is linearly related to the particle's displacement. For clarity, the linearity factor $F(x)$ is used to determine the linear range of the displacement detection:

$$F(x) = \left| \frac{S_F(x) - S_F(0)}{S_F(0)} \right| \qquad (1)$$

where $S_F(x)$ is the slope of the QPD *x*-direction response curve at each point, $S_F(0)$ is the slope of the curve at $x = 0$. According to equation (1), the linear range is defined as the region where $F(x) < 10\%$.

As shown in Fig. 2(a), the *x*-direction signal of QPD shows a roughly linear relationship with the x-axis displacement. When the lateral offset increases, the normalized *x*-direction signal of QPD is less sensitive to the *x*-axis displacement of the particle than the curve shown in $\Delta y = 0$ nm. Figure 2(b) illustrates that the response curves of the *y*-direction signal of QPD presents a parabolic symmetry, and the slope of the curve at $x = 0$ is zero. Ideally, the response of the *y*-direction signal should be zero when the particle moves along the *x*-axis. However, the lateral offset leads to an aggravated response of the *y*-direction signal of QPD.

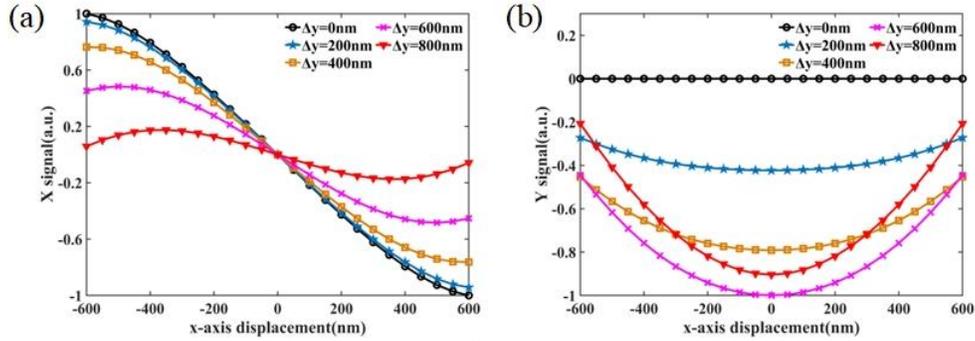

Fig. 2. Simulation of the response curves of QPD signals under various $\Delta y$. (a) The $x$-direction signal of QPD as a function of the $x$-axis displacement. (b) The $y$-direction signal of QPD as a function of the $x$-axis displacement.

Similarly, the axial offset also affects the performance of the displacement detection. Since changing the axial focus offset actually changes the scattering of the particle in the detection beam. It has been expected that the axial offsets have a greater effect on detection sensitivity compared to the lateral offsets. Therefore, in this study, we have searched for the optimal axial offset for the detection sensitivity. Figure 3 shows the response curves of the $x$-direction signal of QPD under various axial offsets. The slope of the curve increases as the axial offsets increases. It can be noted that the best sensitivity was not obtained when there is no axial offset ($\Delta z = 0$ nm), and the relationship between the sensitivity and axial offset is not linear.

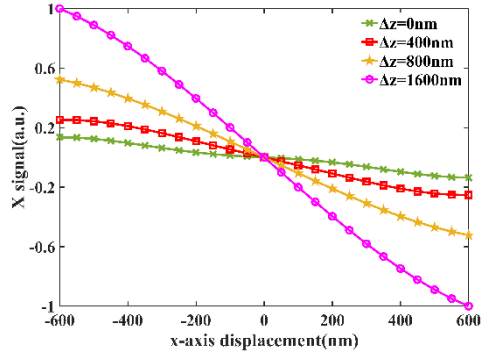

Fig. 3. Simulation of response curves of the $x$-direction signal of QPD under various $\Delta z$.

## 4. Experiment results

### 4.1 Effects of lateral offsets on the displacement detection

When using the detection beam for displacement detection, the low power of the detection beam does not affect the trapping state of the particle, so only the scattering of the particle when it is located inside the detection beam needs to be considered. To verify the effects of the focal offsets on the performance of displacement detection, we scan a particle stuck to a coverslip across the focus of detection beam. Although the displacement detection response using the immobilized particle is different from that of a particle in an optical trap due to effects of light scattering at glass-air interface and spherical aberration of the lens system. This effect is negligible in our experiments, and the variation of detector response at different focal offsets can be intuitively observed by this method.

We deliberately introduce a given lateral or axial offset by the piezo translation stage in experiment. The $x$-direction and $y$-direction signals of QPD are recorded when a silica particle with 10-$\mu$m-diameter is driven through the detection beam focus along the $x$-axis by the piezo translation stage. In this way, we associate the QPD signals to the known particles' displacements, and thus giving the response curves under various offsets.

Figure 4(a) shows the response curves of the *x*-direction signal of QPD with the *x*-axis displacement under various lateral offsets Δ*y* in the experiment. The detection sensitivity and linear range are compared with the simulation results, as shown in Fig. 5. The simulation results and experiment results are denoted as blue solid lines and red triangles, respectively. As shown in Fig. 5(a), the experiment results and simulation results show the same trend that the detection sensitivity decreases with lateral offset Δ*y*. Figure 5(b) shows how the linear range depends on the lateral offset. The results indicate that the linear range decreases with increasing lateral offset, and the experimental results are in agreement with the theoretical results.

Figure 4(b) shows the *y*-direction signal response of QPD when tracking the *x*-axis displacements under various lateral offsets. It can be seen that there is weak response in the *y*-direction signal of QPD in absence of the lateral offset, which is different from the simulation results. The main reasons for the difference include the small asymmetry of the scattered pattern onto the QPD and a difference in sensitivity of the four quadrants of the QPD.

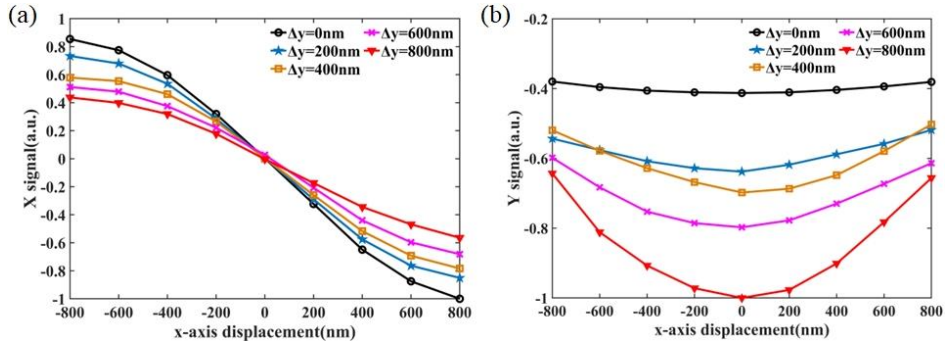

Fig. 4. The response curves of QPD signals under various lateral offsets in experiment. (a)X-direction signals of QPD versus *x*-axis displacements. (b)Y-direction signals of QPD versus *x*-axis displacements.

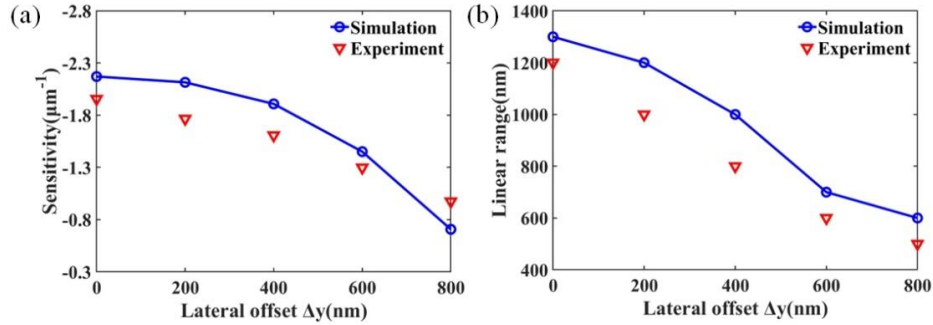

Fig. 5. (a) Detection sensitivity as a function of the lateral offset Δ*y*. (b)Linear range as a function of the lateral offset Δ*y*.

In optical tweezers system, the accuracy of displacement and force detection are often compromised by the crosstalk between the different signals of QPD that inherently exists. The lateral offsets cause the motion information of the *y*-axis to be coupled to the *x*-axis in displacement detection, which will aggravate the crosstalk and thus introduce extra measurement errors. To further quantitatively analyze the influence of the lateral offset on the signal crosstalk, we use the crosstalk coefficient to evaluate the crosstalk between the *x*-direction and *y*-direction signal of QPD resulting from the lateral offsets, which is defined as:

$$C_{xy} = \frac{<V_y - V_0>}{<V_x>} \qquad (2)$$

where $<V_y - V_0>$ and $<V_x>$ are the mean of ($V_y - V_0$) and $V_x$ respectively. $V_y$ is the *y*-direction signal of QPD resulting from the *x*-axis displacement of the particle, and $V_x$ is the *x*-direction signal of QPD resulting from the *x*-axis displacement, $V_0$ is the *y*-direction signal of QPD when there is no lateral offset. The logarithmic form $C_{xy}(dB) = 10\log_{10}(C_{xy})$ is often used here.

According to equation (1), the crosstalk coefficients under various lateral offsets are calculated and compared with the simulation results, as shown in Fig.6. The simulation results and experiment results show that the crosstalk coefficient increases with the lateral offset. This means that as the lateral offset increases, the crosstalk between the *x*-direction and *y*-direction signal of QPD is aggravated.

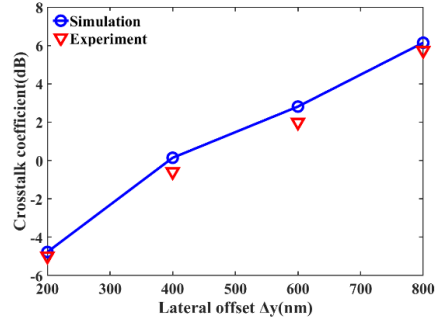

Fig. 6. Experiment and simulation results of the crosstalk coefficient as a function of the lateral offset.

The results presented in Fig. 5 and Fig. 6 indicate that the lateral offset will compromise the sensitivity and linear range of the displacement detection and aggravate the signal crosstalk. The crosstalk shows up as additional power in the power spectral density (PSD) of the *x*-direction and *y*-direction signal when analyzing the PSD of the positional signal. To a certain extent, it limits the performance of the displacement and force detection. This implies that it is imperative to avoid thus extra measurement errors resulting from the lateral offset in the experiment.

### 4.2 Effects of axial offsets on the displacement detection

We take one step further to validate the effects of axial offset on the displacement detection. Figure 7 compares the response curves of the *x*-direction signal of QPD with the *x*-axis displacement under various axial offsets in the experiment. Similar to the simulation results, the slope of the curve increases with the axial offset. When the detection beam focus coincides with the particle centre in the z-axis does not show the best detection sensitivity.

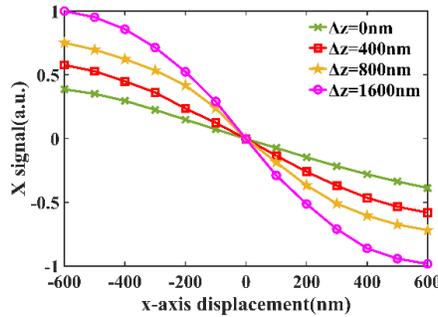

Fig. 7. The response curves of the *x*-direction signals of QPD with the *x*-axis displacement under various axial offsets in experiment.

Figure 8 shows the detection sensitivity as a function of the axial offset in the simulation and experiment. The blue solid line represents the simulation result, and the experiment results are denoted as red triangles. Both the experiment and simulation results show the same trend. As the axial offset $\Delta z$ increasing, the detection sensitivity first increases and at a certain value of axial offset ($\Delta z_{opt}$) reaches to a maximum, after which it starts to decrease. The sensitivity at $\Delta z_{opt}$ is significantly larger than that at $\Delta z = 0$ (no axial offset), and the sensitivity indicates a remarkable axial offset dependence. It is worth mentioning that the presence of an optimal axial offset ($\Delta z_{opt}$) of 1600nm is consistent with the theoretical prediction.

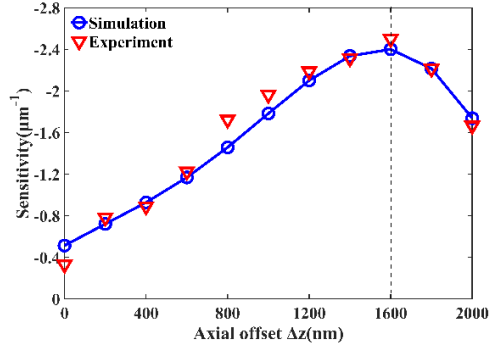

Fig. 8. Detection sensitivity as a function of the axial offset.

The variation in detection sensitivity with axial offset is related to the scattering properties of the particle in the detection beam. Deviating the detection beam focus from the particle centre along the *z*-axis will enhance scattering at small angles. The scattered pattern impinged on the QPD becomes smaller, while the average deflection of detection beam is not substantially affected, thus improving the detection sensitivity. Therefore, one could expect to find an optimal axial offset ($\Delta z_{opt}$) that maximizes the sensitivity of the particles' displacement detection.

## 5. Discussion

We have demonstrated that the performance of the displacement detection can be improve by adjusting the focal offset by both experiment and simulation. The above analysis is for a particle with 5-μm-raduis, but the optimal axial offset required to obtain better sensitivity is different for systems with different parameters. The effects of particle radius *a*, numerical aperture of the condenser $NA_c$ and numerical aperture of the objective $NA_o$ on the optimal axial offset are discussed in this section. The parameters are selected as follows: $\lambda$ = 642 nm, $n_m$ = 1, $NA_c$ = 0.65, $NA_o$ = 0.65, and the remaining parameters remain unchanged except for the one analyzed.

(1) Particle radius *a*

Figure 9 shows the optimal axial offsets $\Delta z_{opt}$ with different particle radius *a*. As shown in Fig. 9, when the particle radius is smaller than the wavelength of the detection laser, varying the axial offsets has little effect on the sensitivity, and the optimal axial offset can be essentially negligible. When the particle radius is larger than the wavelength of the detection laser, the optimal axial offset increases with the increasing particle radius.

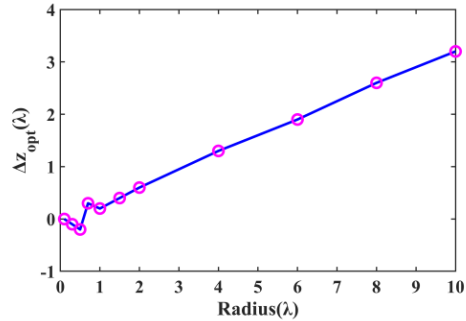

Fig. 9 Optimal axial offsets $\Delta z_{opt}$ with different particle radius *a*

(2) Numerical aperture of the condenser $NA_c$

The effects of the numerical aperture of the condenser $NA_c$ on the optimal axial offset are shown in Fig.10, where $a$ = 8$\lambda$, $NA_o$ = 0.65. It is shown that as $NA_c$ increases, the optimal axial offset $\Delta z_{opt}$ remains essentially constant, and $\Delta z_{opt}$ fluctuates slightly around $2.5\lambda$.

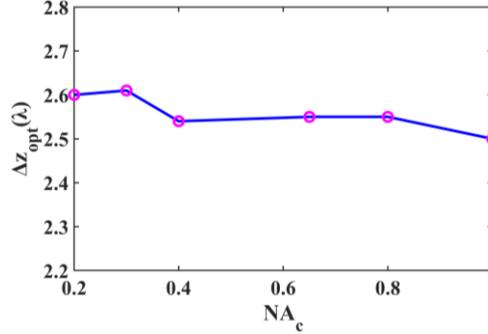

Fig. 10 Optimal axial offsets $\Delta z_{opt}$ with $NA_c$

(3) Numerical aperture of the objective $NA_o$

Figure 11 shows the optimal axial offset $\Delta z_{opt}$ as a function of the numerical aperture of the objective $NA_o$, where $a = 8\lambda$, $NA_c = 0.65$. As can be seen from the Fig. 11, when $NA_o$ is smaller than $NA_c$, the optimal axial offset is negative. When $NA_o$ is comparable to $NA_c$, the optimal axial offset reaches maximum; when $NA_o$ is larger than $NA_c$, the optimal axial offset decreases with increasing $NA_o$.

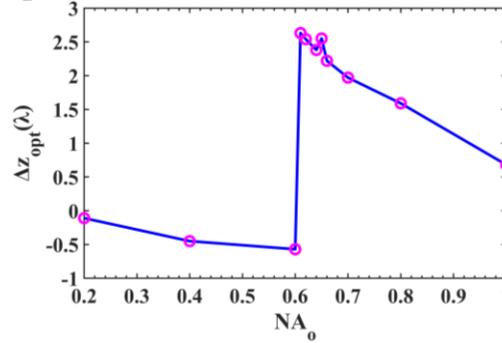

Fig. 11 Optimal axial offsets $\Delta z_{opt}$ with $NA_o$

The above analysis results show that for systems with different parameters, especially for particles with radius larger than the wavelength, it is necessary to obtain the optimal axial offset required for a better sensitivity based on simulation results and experimental verification.

## 6. Conclusion

In this study, we introduce a detection beam into our system for displacement detection. This is useful in experiments where trapping and displacement detection are independent. The sensitivity of displacement detection is measured under various lateral and axial offsets between the detection beam and the particle, and compared to theoretical predictions. The lateral offsets will decrease the detection sensitivity and linear range. On the basis of the theoretical predictions and experimental results, a higher detection sensitivity can be achieved with an optimal axial offset. In addition, the effects of the lateral offsets on crosstalk between the $x$-direction signal and $y$-direction signal of QPD are discussed. The results show that the larger the lateral offset, the larger the crosstalk. The performance of displacement detection can be improved by finely tuning the lateral and axial offsets. Moreover, the effects of particle radius, numerical aperture of the condenser and numerical aperture of the objective on the optimal axial offset have been investigated. The optimal axial offset consistently increases with particle radius. The numerical aperture of the objective can also affect the optimal axial offset, but the numerical aperture of the condenser has a negligible effect on the optimal axial offset.

Based on the analysis of the results, it is found that this displacement detection scheme using a detection beam provides a more flexible working point compared to the conventional displacement detection methods in optical tweezers. It allows for the active modulation of the sensitivity and linear range of the displacement detection. This work provides a theoretical reference and experimental guidance for optimizing the performance of the displacement detection system using a detection beam, which is expected to improve the accuracy of the displacement and force detection performed by the optical tweezers. The improved detection performance will extend the capabilities of optical tweezers in precision measurements.

**Funding.** National Natural Science Foundation of China (61975237, 11904405), Scientific Research Project of National University of Defense Technology (ZK20-14) and Natural Science Foundation of Hunan Province (2021JJ40679).

**Acknowledgments.** The authors wish to thank the anonymous reviewers for their valuable suggestions.

**Disclosures.** The authors declare no conflicts of interest.

**Data availability.** Data underlying the results presented in this paper are not publicly available at this time but may be obtained from the authors upon reasonable request.